# Evolutionary dynamics determines adaptation to inactivation of an essential gene.

**João V. Rodrigues and Eugene I. Shakhnovich***

**Department of Chemistry and Chemical Biology, Harvard University, 12 Oxford Street, Cambridge, MA 02138, USA**

*\* Corresponding author*

email  EIS:  shakhnovich@chemistry.harvard.edu

**Abstract**

Genetic inactivation of essential genes creates an evolutionary scenario distinct from escape from drug inhibition, but the mechanisms of microbe adaptations in such cases remain unknown. Here we inactivate *E. coli* dihydrofolate reductase (DHFR) by introducing D27G,N,F chromosomal mutations in a key catalytic residue with subsequent adaptation by serial dilutions. The partial reversal G27->C occurred in three evolutionary trajectories. Conversely, in one trajectory for D27G and in all trajectories for D27F,N strains adapted to grow at very low supplement folAmix concentrations but did not escape entirely from supplement auxotrophy. Major global shifts in metabolome and proteome occurred upon DHFR inactivation, which were partially reversed in adapted strains. Loss of function mutations in two genes, *thyA* and *deoB*, ensured adaptation to low folAmix by rerouting the 2-Deoxy-D-ribose-phosphate metabolism from glycolysis towards synthesis of dTMP. Multiple evolutionary pathways of adaptation to low folAmix converge to highly accessible yet suboptimal fitness peak.

## Introduction

The ability of microorganisms to adapt to different environments is crucial for their long-term survival and particularly important in the development of antibiotic resistance. Bacterial metabolism is often equipped with alternative pathways specialized in the utilization of diverse nutrients, which provides microbes an exceptional ability to thrive in various environments and under stress from antibiotic treatment. In addition, multi-layers of regulation that global control of cellular functions ensures both robustness, against fast fluctuating conditions, and flexibility, through slower transcriptional responses to alternating nutrients(Bennett et al., 2008; Buescher et al., 2012; Millard et al., 2017). When important cellular functions are inactivated, e.g. by genetic mutations, a severe disruption of cellular networks can occur, which poses a major adaptive challenge to cells. Recent studies investigated evolution of *E. coli* upon inactivation of non-essential enzymes of carbon metabolism that lead to re-wiring through less efficient pathways (Krusemann et al., 2018; Long et al., 2018; McCloskey et al., 2018a, b, c, d). These studies highlight a crucial interplay between regulatory responses and imbalances in metabolite concentrations resulting from gene knockouts, which can be subsequently corrected by mutations elsewhere. Importantly all these studies involved gene-knockout so that adaptation by reversal to WT genotype was not possible.

Less is known about the adaptation mechanisms that follow inactivation of unique cellular processes that are deemed indispensable in microbes. The genes that perform such functions, often classified as essential, are believed to face higher selective pressure and thus evolve slower (Luo et al., 2015). It is thus expected that disruption of essential genes can create a major barrier to the adaptability of bacteria. Nevertheless, a comprehensive study involving knock outs of >1000 genes classified as essential in yeast has shown that a small percentage of mutants could recover viability after laboratory evolution (Liu et al., 2015), revealing that essentiality can, in some cases, be overcome through adaptation. Loss of essential biosynthetic genes have also been observed to occur naturally, under evolutionary conditions where pathway products are provided externally (D'Souza and Kost, 2016), highlighting a context-dependent nature of essentiality. Nevertheless, while adaptation upon inactivation of essential genes in microbes has been demonstrated, its mechanisms remain unknown.

Here we study evolutionary adaptation upon functional inactivation of an essential *E. coli* enzyme dihydrofolate reductase (DHFR). Past efforts to link fitness effects of chromosomal variation in the *folA* locus encoding DHFR and their biophysical effects on DHFR protein allowed us to develop an accurate quantitative biophysical model of DHFR fitness landscape (Bershtein et al., 2015a; Bershtein et al., 2013; Bershtein et al., 2012; Bershtein et al., 2015b; Rodrigues et al., 2016). The availability of a clear genotype-phenotype relationship makes DHFR a unique model to study the dynamics and outcomes of evolution to recover from its inactivation by point mutations. In contrast to previous studies that used gene knockouts here we inactivate the chromosomal *E. coli* DHFR by introducing mutations in the *folA* locus at a key catalytic residue in position 27, generating strains that express inactive DHFR protein but are viable only with external metabolic compensation, allowing the mutants to adapt to lack of DHFR function by decreasing supplement concentration. The advantage of this approach is that it presents cells with an obvious evolutionary "solution" of reverting the mutant back to WT-variant without massive rewiring that could lead to potentially lesser fit variants. However, an actual outcome may depend on evolutionary dynamics which could revert to WT or other form of active DHFR (higher fitness peak) or converge to a potentially more accessible solution of rewiring to a less efficient metabolic pathways that do not require DHFR function. Therefore, this setup allows us to assess relative roles of height and accessibility of fitness peaks in determining the outcome of evolutionary dynamics.

We show that, depending on starting DHFR variant, partial reversion of DHFR phenotype may indeed occur. However, adaptation to low concentration of external metabolites through metabolic rewiring is the prevalent evolutionary solution due to the availability of a greater number of trajectories leading to consecutive gene inactivation events in two key loci, *thyA* and *deoB*. Using omics analysis, we observe global perturbations in metabolites and proteins levels occurring due to DHFR inactivation and upon adaptation, highlighting a key role of regulatory circuits in directing evolution. Finally, we show how adaptation to loss of drug target generates highly resistant strains, and that one important evolutionary solution found here, inactivation of thyA gene, can be also found in clinical isolates of resistant strains of *S. aureus* (Chatterjee et al., 2008; Kriegeskorte et al., 2014) and *H. influenza* (Rodriguez-Arce et al., 2017).

**Results**

DHFR is encoded by the *folA* gene and is essential for the biosynthesis of purine, pyrimidine and glycine. We sought to inactivate *folA* in *E. coli* by introducing mutations in the key catalytic residue D27 (Figure 1A).

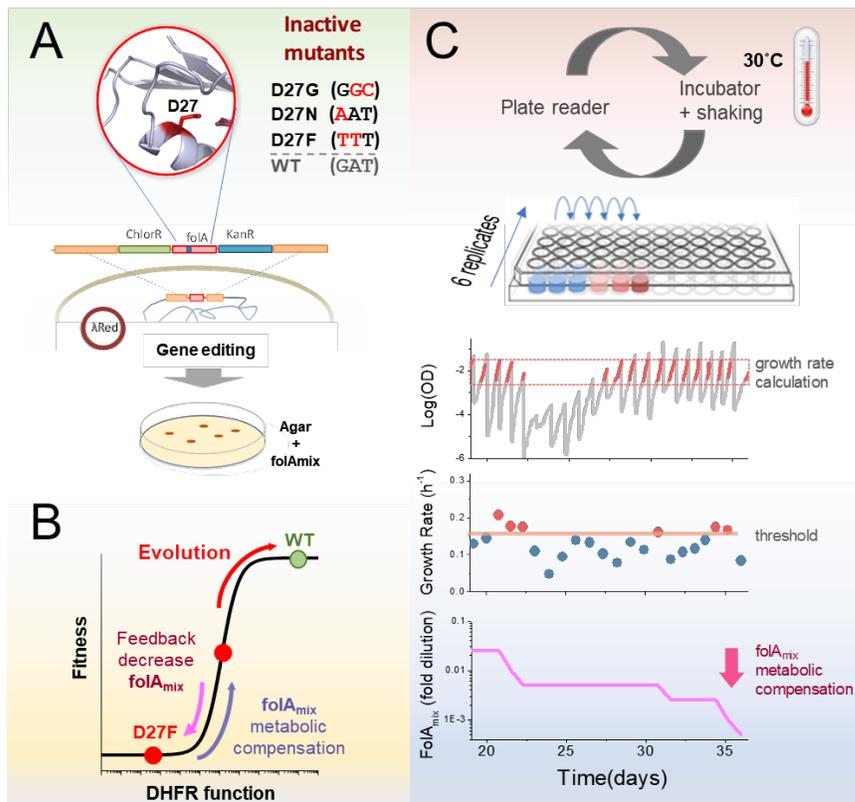

**Figure 1 – Automated experimental evolution of inactive DHFR mutants.** *A) Mutations in key D27 catalytic residue of DHFR were introduced in E. coli chromosome, together with flanking antibiotic resistant markers, by lambda red recombination and strains devoid of DHFR function were selected with antibiotics in folAmix-containing agar plates. B) D27 mutants are lethal but fitness can be partially rescued by metabolic compensation with folAmix. Adaptative changes that increase fitness can be counterbalanced by decreasing the concentration of folAmix in the growth medium, forcing the cells to evolve without DHFR function. C) Experimental evolution using automated liquid handling. Six replicates of cell cultures are placed in the first column of a 96well plate and incubated at 30˚C with shaking and the optical density is measured every 30 min. The cultures grown until the average OD reaches a threshold (0.3), that was defined to avoid nutrient limitation and consequent shift into the stationary phase. At this point cultures are diluted simultaneously into the wells of the adjacent column, to a common starting OD (0.01), and the cycle is repeated. The growth rate is calculated for every culture and whenever it exceeds a defined threshold the amount of folAmix is independently decreased for each culture in the subsequent dilution.*

However, such approach poses a methodological challenge since, by definition, the knockout of essential genes results in lethality or infertility. However, *folA* knockout mutants are conditionally lethal if grown in a culture medium supplemented with a mixture of metabolites comprising thymine, glycine, methionine, inosine and adenine (folAmix) (Howell et al., 1988; Kwon et al., 2010). We envisaged that inactivated DHFR mutants growing in the presence of folAmix could be progressively challenged to adapt to ever decreasing concentrations of this supplement in the growth medium (Figure 1B). The next step was to develop a protocol that allowed controllable decrease of supplement concentration in growth medium in response to the fitness status of the cells. This was achieved by implementing a fully automated serial passaging scheme as depicted in Figure 1C-D using a Tecan liquid handling robot. In this setup, the growth rate of replicate cultures is monitored by periodic OD readings, and the concentration of folAmix is adjusted downward in each serial passage when cultures exceed a defined growth rate threshold. This feedback control loop ensures mutant strains are continuously challenged to grow at sub-optimal conditions, sustaining a strong selective pressure on the loss of DHFR function. This approach combines the medium-throughput capabilities of plate-based serial dilution methods (currently, up to 32 independent replicate trajectories can be evolved in parallel), and both real-time monitoring of fitness status and control of growth conditions featured in morbidostat setups (Toprak et al., 2012; Toprak et al., 2013), which allows cells to continuously experience exponential growth and sustained selection pressure.

*D27 mutations confer growth defects*

We selected single or double nucleotide base mutations to replace a key catalytic residue Asp 27 either for Asn, Gly or Phe; the presence of a carboxylate side chain in position 27 is strictly conserved among DHFR orthologues. Purified D27 mutant proteins were characterized (Table 1) and the catalytic activity measurements confirm the lack of significant DHFR function in these variants; catalytic efficiencies ($k_{cat}/K_M$) are several orders of magnitude lower than WT. However, thermal denaturation data obtained for the D27 mutant proteins indicates that their stability is mostly unaffected (or even significantly increased in the case of D27F mutant (Tian et al., 2015b)) showing that, despite the lack of catalytic activity, these proteins retain the ability to fully fold. Importantly, this provides a pathway for restoration of DHFR function over the course of evolution, either by revertant mutations at the D27 locus or, potentially, compensatory mutations elsewhere in the protein. D27 mutant strains were generated by lambda red recombination (Bershtein et al., 2012) and plated in folAmix-containing agar media, however, growth was only visible after 48h and the colonies formed were miniscule. When DHFR function was assayed in cell lysates of D27 mutant strains, no DHFR activity could be detected, confirming that these mutations inactivate DHFR. As expected, growth experiments showed that D27 mutants grow only at high concentrations of folAmix (Figure 2).

**Table 1- In vitro properties of DHFR mutant proteins[a]**

| | WT | D27F | D27N | D27G | D27C |
|---|---|---|---|---|---|
| $k_{cat}$ (s$^{-1}$) | $1.4 \times 10^{1}$ | $4 \times 10^{-4}$ | $1 \times 10^{-2}$ | $1 \times 10^{-2}$ | $2.1 \times 10^{0}$ |
| $K_M$ (μM) | $8.7 \times 10^{-1}$ | $1 \times 10^{2}$ | $1 \times 10^{1}$ | $3 \times 10^{1}$ | $8 \times 10^{1}$ |
| $k_{cat}/K_M$ (s$^{-1}$ μM$^{-1}$) | $1.6 \times 10^{1}$ | $4 \times 10^{-6}$ | $7 \times 10^{-4}$ | $4 \times 10^{-4}$ | $3 \times 10^{-2}$ |
| $k_{cat}/K_M$ relative to WT | 1 | $2 \times 10^{-7}$ | $4 \times 10^{-5}$ | $3 \times 10^{-5}$ | $2 \times 10^{-3}$ |
| $K_i$ (nM) | 0.94 | ND | ND | ND | $1.7 \times 10^{4}$ |
| $\Delta T_m$ ($^{0}$C) [a] | 0 | +7.6 [b] | +1.0 | -0.3 | -1.5 |
| bis-ANS | 1 | 1.0 | 1.2 | 2.4 | 1.3 |

[a] Kinetic properties for dihydrofolate reductase catalytic activity ($k_{cat}$ = enzymatic turnover number, $K_M$ = Michaelis-Menten constant, $k_{cat}/K_M$ = catalytic efficiency, $K_i$ = inhibition constant for trimethoprim) and protein stability properties (ΔTm = difference in melting temperature of folding with respect to wild type, bis-ANS = relative fluorescence from binding of bis-ANS to molten-globule intermediates with respect to wild type protein.)

[b] from Ref (Tian et al., 2015b)

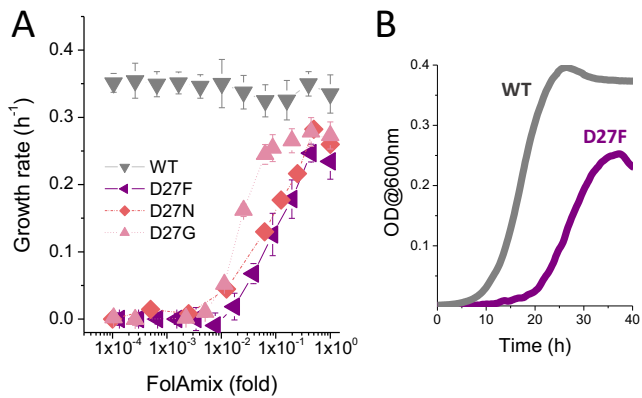

**Figure 2** – *D27 mutants require folAmix to grow. A) The growth rates of each D27 mutant and wild type were measured at various dilutions of folAmix, with respect to initial composition (adenine 20 µg/mL, inosine 80 µg/mL, thymine 200 µg/mL, methionine 20 µg/mL and glycine 20 µg/mL). Cultures were grown in M9-minimal medium at 30˚C, and absorbance was monitored at 600nm. B) Comparison of representative growth curves of wild type and D27F mutant obtained in the presence of 1x folAmix*

Interestingly, the D27G mutant shows slightly better growth at lower concentrations of folAmix than D27F and D27N. Given that the catalytic efficiency of D27G and D27N mutants is comparable, and extremely low, it is very likely that the difference in growth originates from a potential acquisition of a slightly advantageous mutation elsewhere upon genetic manipulation to introduce D27G mutation in the folA locus (see Discussion). We tested D27 mutants for growth in the presence of individual components of folAmix and their different combinations and found that only thymine was essential, although growth with thymine alone is slower compared to growth with all folAmix components (Supplementary Figure S1). The growth rates measured for all D27 mutants at high folAmix fall in the range 70-80% of WT and lag times were typically 1.5-2 fold longer (Figure 2 B). Overall, weaker growth and small colony phenotype show that D27 mutants are severely compromised even in presence of folAmix. Using the evolutionary scheme described previously, we allowed six replicates of each mutant strain to evolve in parallel for about 50 days.

*Evolution of D27G mutant*

Cultures of D27G mutant strains were first grown in the presence of folAmix diluted 0.1 fold with respect to the initially defined folAmix composition. Throughout the course of the experiment the concentration of supplement mixture further decreased, as imposed by the feed-back control loop, in response to increasing fitness of the mutant strains (Figure 3A).

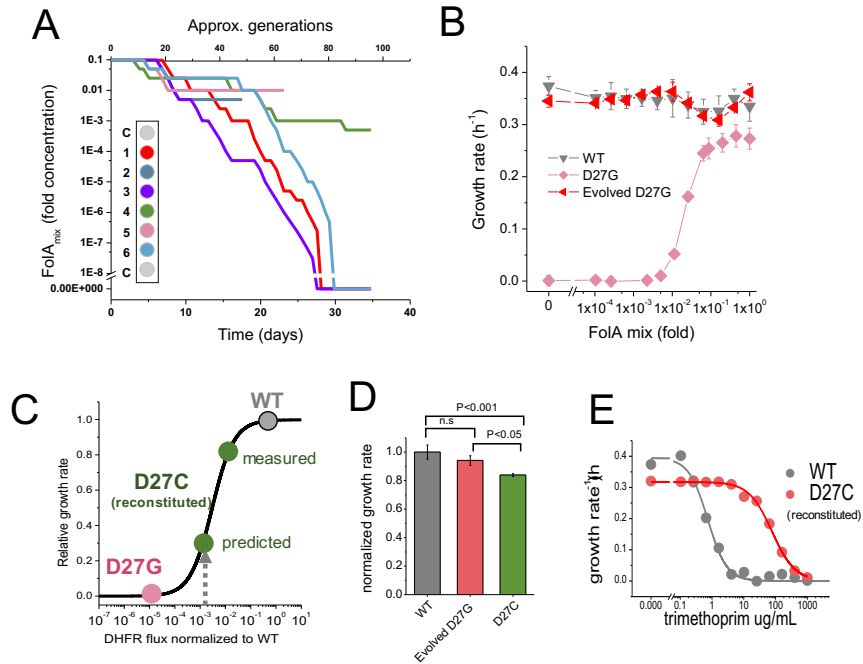

**Figure 3- Phenotype-reverting G27C mutation emerges upon evolution of D27G strain.** *A) Evolutionary profiles of the six trajectories in D27G adaptation showing the time dependent changes in folAmix concentration necessary to sustain growth. Each trajectory is colored according to its position in the column of the 96-well plate, as represented in the scheme (C = wells with growth medium only). B) Evolved D27G strain does not require folAmix to grow. The growth rates determined at different concentrations of folAmix are compared for wild type, naïve and evolved D27G strains. Data are represented as mean ± SD. C) Biophysical properties determined in vitro for the D27C DHFR were used to predict growth rate based on the DHFR fitness(Rodrigues et al., 2016). The resulting prediction (30% of the wild type value) is in fair agreement with the experimental value measured for a BW25113 strain in which D27C mutation was introduced in the wild type folA gene (D27C reconstituted). D) Comparison of growth rates of wild type, evolved D27G and reconstituted D27C strain. E) Dose-dependent growth inhibition by trimethoprim determined for wild type and reconstituted D27C mutant. Solid lines are fits with logistic equation, from which IC50 was determined to be 0.7 ± 0.1 µg/mL for wild type and 77 ± 6 µg/mL for D27C mutant.*

The change in folA mix concentration over time reflects the effect of adaptation. Three trajectories (1,3 and 6) stand out by reaching the ability to grow in the complete absence of folAmix. Trajectories 2 and 5 reached a point where growth dropped below detection limit, and ultimately could not be recovered, whereas trajectory 4 had adapted to grow at low concentrations of folAmix. We hypothesized that mutations in folA locus could have restored DHFR activity in the cultures where reversion of phenotype was observed. Accordingly, Sanger sequencing analysis of this region revealed that in all these three trajectories residue Gly27 had mutated to cysteine. This finding was surprising because alignment of multiple known DHFR sequences shows that cysteine does not occur naturally in position 27; only aspartate or glutamate are observed in this locus. To assess whether Cys27 mutant is functional, this variant was purified and characterized in vitro, and its catalytic properties are compared with wild type and other mutants in Table 1. Although stability-wise D27C mutant is similar to wild type protein, it shows much weaker catalytic efficiency, mostly in terms of $K_M$ retaining only about 0.1% $k_{cat}/K_M$ of wild type. To assess if this residual catalytic function is sufficient to explain the reversion of phenotype in the evolved strain we first predicted fitness based on a previously established model (Rodrigues et al., 2016). Taking as input the vitro biophysical properties of purified D27C mutant, namely kcat/KM and bis-ANS fluorescence values, the model predicts a growth rate of about 30% of the wild type strain. To check the validity of this prediction, the mutation D27C was reconstituted in the wild type background by lambda red recombination and cells were plated in folAmix-containing agar plates to lift any selective pressure for DHFR activity. This strain was able to grow in the absence of folAmix supplementation, and the measured growth rate was 85% of the wild type, which is in fair agreement with the in vitro-based prediction which does not take into account a possibility of upregulation in response to DHFR deficiency (Bershtein et al., 2015a) (see Supporting Information for discussion) (Figure 3B-D). This analysis shows that D27C mutation alone explains most fitness recovery of the evolved strains. Of note is also the fact that significant loss in catalytic efficiency in D27C mutant pays-off with a dramatic 4 orders of magnitude increase in the inhibition constant for trimethoprim. (Figure 3E) This is in line with previous observations that active site mutations compromise catalytic function but also disrupt pocket interactions with the drug, resulting in high levels of resistance (Bader et al., 2006; Rodrigues et al., 2016).

*Evolution of D27F and D27N strains*

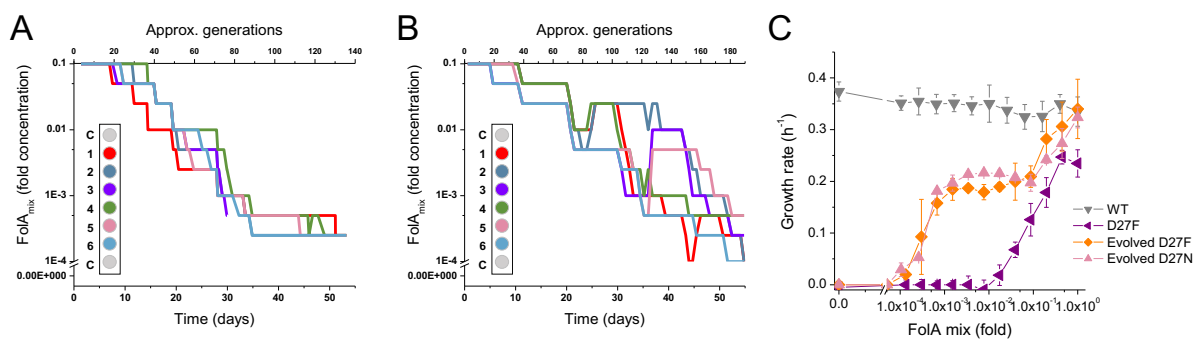

**Figure 4 – D27F and D27N mutant evolved to grow at low folAmix concentrations.** *Evolutionary profiles of each trajectory in A) D27F and B) D27N adaptation to loss of DHFR function. C) Comparison of growth dependence on folAmix concentration determined for wild type, both naïve and evolved D27F and evolved D27N strains, see also Figures S1 and S2. Data are represented as mean ± SD.*

Evolutionary trajectories of D27N and D27F mutants represented in Figure 4 A-B show a marked decrease in folAmix necessary to sustain growth, reaching concentrations nearly three orders of magnitude lower than initially required for naïve strains. Adaptation to grow at low folAmix concentrations was verified by showing that growth curves of individual colonies isolated from evolved strains are similar to the ones obtained in evolutionary dynamics (Figure 4C). However, these strains cannot grow in the complete absence of folAmix. This result indicates that, unlike some trajectories in D27G evolution, adaptation of D27N and D27F strains did not involve reversion of the *folA*- phenotype. Accordingly, no mutations were found in *folA* locus in evolved D27N and D27F strains. Other mechanisms must therefore be responsible for the ability to grow at low folAmix concentrations. The growth rate of the evolved strain measured at high concentrations of folAmix reached values comparable to wild type, showing a clear fitness increase with respect to naïve strain. However, at lower concentrations of folAmix the growth rate of evolved strain becomes markedly reduced to nearly half of the wild type value and appears to plateau in an intermediate range to folAmix concentration. We observed a similar plateau when growth was measured at different concentrations of thymine alone (Supplementary Figure S1), from which we concluded that at low folAmix concentrations only thymine is being essential to sustain growth whereas other components are probably too diluted to have a positive impact in fitness. We also evaluated sensitivity to trimethoprim for both naïve and evolved D27F strains. Not surprisingly, in the absence of a functional DHFR and in the presence of folAmix D27F mutant strains are extremely resistant to trimethoprim (Supplementary Figure S2A). We noted, however, that while no decrease in growth was evident in naïve D27F strain up to 1000 µg/mL, the growth rate of the evolved strain dropped abruptly above 500 µg/mL trimethoprim, suggesting that the drug is acting on an unknown target in the cell which is essential in the evolved but not the parent strain. We then tested how the resistance of evolved D27F strain would compare with wild type at very low concentrations folAmix (Supplementary Figure S2B). We found that in those conditions IC50 for the wild type is similar to that measured in the absence of supplement, yet the evolved D27F still shows an extremely high resistance to trimethoprim (IC50=656±78 µg/mL).

*D27 mutation causes major metabolic and proteomic changes.*

The previous observation that D27 mutant strains show severe growth defects suggests that DHFR inactivation imposes considerable homeostatic imbalance. We focused on the strain D27F to carry out a detailed characterization of the systems-level effects of DHFR inactivation. To that end we carried out high throughput proteomics analysis of D27F mutant strains using LC/MS TMT approach as described earlier (Bershtein et al., 2015a). The method based on differential labeling provides abundances of proteins in the proteome relative to a reference strain which in our case was WT (Bershtein et al., 2015a). We computed Z-scores of the log of relative (to wild type as reference) abundance according to the following equation:

$$z_i^{strain/ref} = \frac{Y_i^{strain/ref} - \langle Y^{strain/ref} \rangle}{\sigma_Y^{strain/ref}}, \text{ eq. 1}$$

where index *i* refers to gene, $Y_i^{strain/ref} = Log_{10}\left(\frac{A_i^{strain}}{A_i^{ref}}\right)$ where $A_i^{strain}$ and $A_i^{ref}$ are the gene i abundances obtained for the mutant and reference (wild type) strains, respectively, $\langle Y^{strain/ref} \rangle$ denotes a quantity $Y_i^{strain/ref}$ averaged over all genes for a given strain, and $\sigma_Y^{strain/ref}$ is the standard deviation of $Y^{strain/ref}$. Then, we performed a Student t-test to determine which groups of genes had statistically significant variation of protein levels in naïve D27F strain, with respect to wild type, following the functional and regulatory classification of genes into groups by Sangurdekar et al (Sangurdekar et al., 2011). Numerous processes were significantly altered, reflecting broad genome-wide effects of DHFR inactivation (Supplementary Figure S3). Several processes were downregulated in naïve D27F strain, from energy metabolism (aerobic respiration, TCA cycle), to the metabolism of nitrogen, several amino acids, pyrimidines and lipopolysaccharides. On the other end, we found significant upregulation of stress responses, peptidoglycan recycling and salvage of guanine and xanthine. We then performed both targeted and untargeted metabolomic analysis of naïve D27F mutant strain and wild type (see experimental details) to characterize significant changes at the level of metabolites. Likewise, Z-scores for metabolite levels were computed for naïve D27F mutant, with respect to wild type. The metabolites with the highest absolute Z-scores (>1.96) were selected for an enrichment test using MBRole online software (Lopez-Ibanez et al., 2016) to identify pathways in which altered metabolites are overrepresented. The analysis revealed that the metabolism of purines, pyrimidines, beta-alanine, histidine and sulfur were the most significantly changed in naïve D27F mutant comparatively to wild type (Supplementary Figure S4). A detailed scheme representing the changes in metabolites and proteins levels of nucleotide synthesis pathways is shown in Figure 5A.

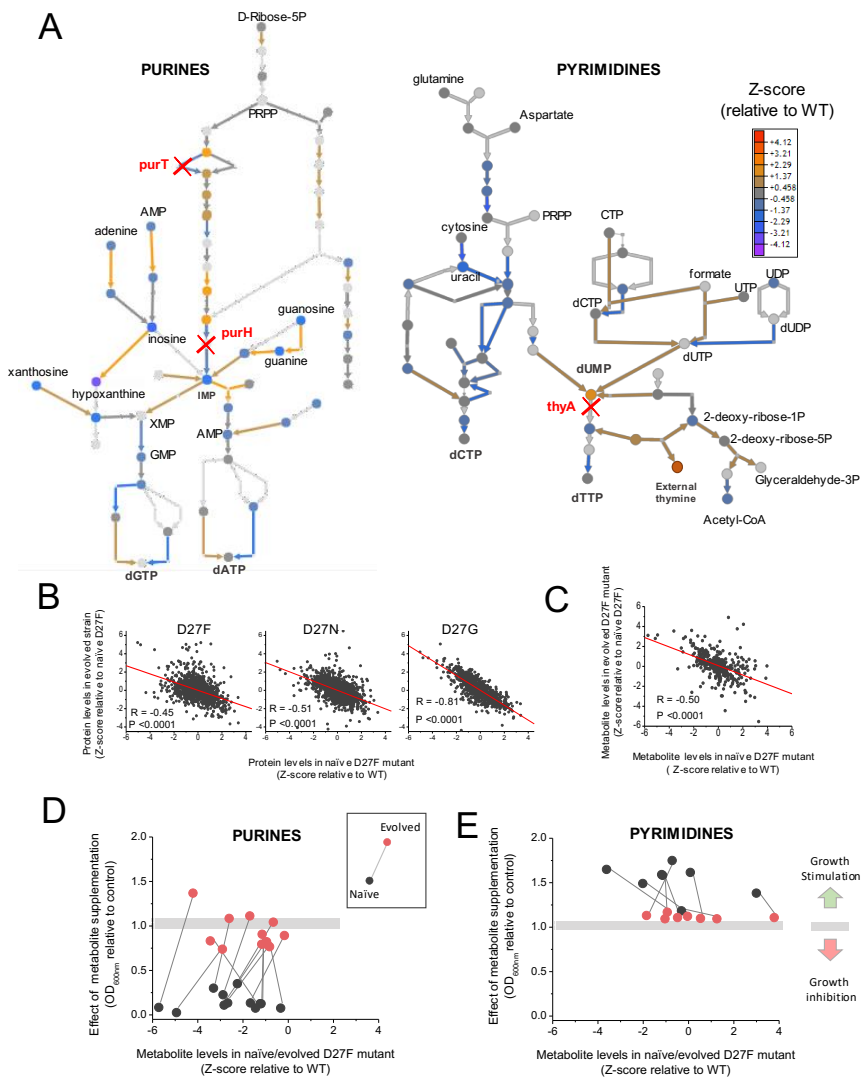

**Figure 5 – D27 mutation causes major metabolic and proteomic changes.** *A) Schematic representation of purine and pyrimidine biosynthetic pathways depicting changes in the levels of metabolites and proteins for the naïve D27F mutant (Z-scores, relative to wild type). Metabolites are represented by circles and arrows represent enzymatic reactions color-coded by the levels of associated proteins (see also Figures S3 and S4). Light gray shading indicates that no data is available. Metabolites that are downstream of the enzymatic reactions requiring reduced folates (ThyA, PurH and PurT) are strongly depleted in naïve D27F mutant, whereas those metabolites immediately upstream to those reactions are increased in respect to wild type. B) Proteomics changes upon evolution partially revert the effect of DHFR inactivation (see also Figures S5 and S6). Comparison of the changes in protein levels obtained for evolved D27F, D27N and D27G strains (Z-scores, relative to naïve D27F) with those measured for naïve D27F (Z-scores, relative*

*to wild type). C) Comparison of the changes in metabolite levels measured for evolved D27F strain (Z-scores, relative to naïve D27F) with those measured for naïve D27F (Z-scores, relative with wild type). D-E) Inhibitory or stimulatory effect of purine/pyrimidine metabolites supplementation on the growth of the naïve D27F strain is attenuated in the evolved D27F strain. The intracellular levels of various purine and pyrimidine species including bases, (deoxy)nucleosides and (deoxy)nucleotides, were determined by targeted metabolomics for naïve (black circles) and evolved (red circles) D27F strains, represented as Z-scores relative to wild type (see also supplemental Figure S5). The effect of supplementation with individual metabolites on growth was determined by growth measurements of naïve and evolved D27F mutant strains in M9 minimal medium supplemented by 1.6 mM thymine in combination with one of each tested metabolites (1 mM final concentration). Relative growth represents the maximum OD obtained for each metabolite normalized to the value measured in the presence of thymine alone. Although most purine metabolites are strongly depleted in naïve D27F mutant, supplementation of the culture medium with these compounds has a strong inhibitory effect on growth. Contrastingly, pyrimidines have a stimulatory effect on growth. In evolved D27F strain, the supplementation with purines and pyrimidines has only a marginal effect on growth*

Not surprisingly, we observe build-up of metabolites upstream the reactions that require reduced folate cofactors (purT, purH, and thyA), whereas metabolites downstream of those reactions are found to be strongly depleted.

*Evolved D27 strains partially revert the omics effects of DHFR inactivation*

Next, we carried out same LC/MS TMT proteomics analysis of evolved strains (D27F, D27N and D27G) to help identify systems-level changes emerging from adaptation to lack of DHFR function. To get a glimpse of global proteomics changes we applied PCA analysis which revealed that both wild type and evolved D27G occupy the same quadrant in the space of two principal components, whereas D27 mutants that are folAmix dependent cluster more closely in another quadrant (Supplementary Figure S5). We then computed the Z-scores for the proteomic changes in evolved strains with respect to naïve D27F ($Z^{D27\_evo/D27F\_naïve}$), to assess the effect of evolution on the proteomic levels, and plotted these values against $Z^{D27F\_naïve/WT}$ to compare with the initial changes caused by D27F mutation. A clear anticorrelation was observed for all mutants (Figure 5B), being the strongest for evolved D27G strain. These results indicate that adaptation leads to proteomic changes that generally oppose the immediate effects caused by DHFR inactivation. In the case of D27G, the strong global proteomic shift towards wild type levels is somewhat expected since in the evolved strain, the DHFR activity is partly restored by G27->C mutation. To better characterize the proteomic changes that were specific to evolution of folAmix dependence we performed a functional and regulatory classification analysis to identify which groups of genes were significantly altered in both evolved D27F and D27N, with respect to naïve D27F strains. We found an upregulation of genes involved in glycolysis, fermentation, sugar alcohol degradation and response to low pH, suggesting a metabolic shift towards mixed acid fermentation as a result of adaptation. On the other hand, the processes of DNA restriction/methylation and D-ribose uptake were significantly down regulated with respect to naïve D27F in both evolved D27F and D27N strains. Next, we focused on evolved D27F strain to perform a detailed metabolomic analysis and characterize the changes occurring at the level of metabolites. We computed Z-scores for metabolite changes with respect to naïve D27F ($Z^{D27\_evo/D27F\_naïve}$) and we found significant anticorrelation with Z-scores obtained for naïve D27F ($Z^{D27F\_naïve/WT}$) (Figure 5C), indicating that metabolite concentrations in evolved strain generally change to partially recover wild type levels, as observed with proteomics. We found that the pathways significantly enriched in metabolites with the highest $Z^{D27\_evo/D27F\_naïve}$ scores (>1.96) in evolved D27F were the same as those that had the most significant changes in naïve D27F, with respect to wild type (Supplementary Figure S6). Overall these results show that adaptation to low folAmix concentrations is accompanied by an overall shift in the abundance of metabolites and proteins to partially revert the system-level changes that are caused by DHFR inactivation.

*Regulatory responses are altered in evolved D27F strain*

Metabolites of purine and pyrimidine biosynthetic pathways were among the most strongly depleted in naïve D27F strain and showed highest increase upon evolution (Supplementary Figure S4). We reasoned that these depleted metabolites could be limiting growth of the D27 mutants, and that supplementing the culture medium with these metabolites would improve growth. Surprisingly, supplementation with purines strongly inhibited growth of the naïve D27F strain, whereas pyrimidines addition had a stimulatory effect (Figure 5D, and Supplementary Figure S7). The detrimental effect of purine supplementation suggests that drop in purine metabolites in naïve D27F was a regulatory response to high levels of the purines supplied by the culture medium. In the case of evolved D27F strain, the effect of supplementation of individual nucleotides on growth was comparably weaker or non-existent, indicating that the inhibitory/stimulatory effects were relaxed upon adaptation. It is possible that the profound metabolic and proteomic changes upon DHFR inactivation amount to cells switching their proteomes and metabolomes to a novel stable state. In these conditions, the "programmed" responses that are triggered by high external concentrations of metabolites might actually worsen the fitness of the cell, instead of providing the beneficial advantage that they have evolved for. On the other hand, after evolution, even though metabolites and proteins levels are still quite distinct from wild type values, the regulatory network appears to be better adapted to that new metabolic state. Adaptation can either result from changes in the regulatory network itself, due to mutations in regulatory genes, or from re-wiring of metabolic pathways that establish a different metabolic state for which existing regulatory networks provide more optimal responses.

*Whole genome sequencing*

Whole genome sequence (WGS) analysis was performed to identify the genetic basis for the systems-wide changes observed in evolved strains. In addition, to characterize the dynamics of mutation fixation, individual colonies from intermediate evolutionary time points of a selected D27F trajectory were also sequenced.

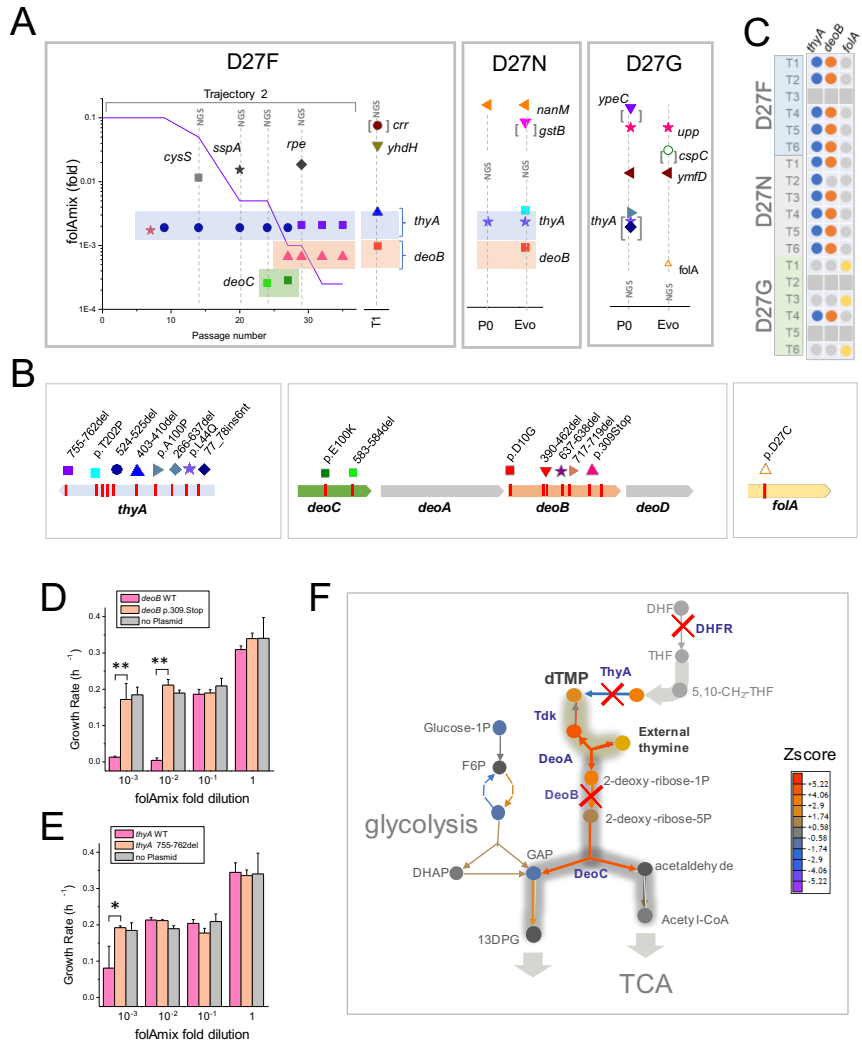

**Figure 6 – Loss of function mutations in two genes, *thyA* and *deoB*, lead to adaptation to low folAmix concentrations.** A) Mutations identified in naïve and evolved D27 mutant strains. Vertical lines identified with NGS represent mutations identified by whole genome sequencing whereas the remaining cases were identified by Sanger sequencing. Details are presented for trajectory 2 of D27F evolution and include the mutations identified at various passages and the folAmix profile obtained for that trajectory. B) Mutations identified in the most relevant genes. C) Mapping *thyA*, *deoB* and *folA* mutations found among evolved strains from all trajectories (1 to 6) of each D27 mutant. Colored dots represent the presence of mutations, whereas gray dots represent genes that were not mutated in evolved strains. Trajectories that ceased growth prior to the end of the experiment were not sequenced (gray squares). D-E) Expression of wild type DeoB and ThyA in evolved D27F strain reverts phenotype to high folAmix requirement. Evolved D27F strain, with background mutations *deoB* p.D10G and *thyA* 403-410del, was transformed with pTRC-tetR

*plasmid coding D) wild type or mutant deoB and E) wild type or mutant thyA, under the control of tetR repressor. Evolved D27F strain without plasmid is also represented as control. Cells were grown in M9 minimal medium and growth rates were measured from periodic OD measurements. Data are represented as mean ± SD. \* P<0.01, \*\* P<0.001. F) Inactivation in deoB gene prevents 2-deoxy-D-ribose-1-phosphate from being diverted into energy production via glycolysis and tricarboxylic acid cycle. Changes in the levels of metabolites (depicted as circles) and proteins (depicted as arrows) in the evolved D27F strain (Z-scores, relative to naïve D27F mutant) are represented. The genes involved in 2-deoxy-D-ribose-1-phosphate degradation are marginally increased in evolved D27F, with respect to naïve strain (average Z-score = 0.445, P=0.0817, see Figure S6). However, deoB inactivation saves 2-deoxy-D-ribose-1-phospahte that is required in dTMP production, instead of being utilized for energy production in glycolysis, which is significantly upregulated in evolved D27F, with respect to naïve D27F (average Z-score= 0.593, P= 0.015*

Figure 6 A-B summarizes the WGS results obtained for evolved and naïve strains from each D27 mutant. We found that D27N and D27G strains, but not D27F, had additional background mutations that must have been acquired prior to starting the evolution experiments; although all D27 strains were constructed from the same parent wild type strain, we could not prevent the appearance of mutations at any given stage prior to the evolution experiments. It is possible that these mutations may provide some fitness advantage with respect to "pure" D27F strain (see below and discussion). Comparison between evolved and naïve D27G strains reveals that no other mutation was fixed upon evolution besides the one nucleotide change at the D27 locus of the *folA* gene, corresponding to the Gly->Cys substitution described earlier. This is fully consistent with the earlier conclusion that D27C mutation alone explains the fitness recovery observed in D27G evolution. From the analysis of evolved D27F and D27N we can identify two common hotspots for mutations, the genes *thyA* and *deoB*, encoding thymidylate synthase and phosphopentomutase, respectively. For that reason, the sequence of these two loci were determined by Sanger analysis for all other trajectories that evolved to grow at low folAmix concentrations, including trajectory 4 in D27G evolution. Strikingly, in a total of 12 trajectories all were found to have mutations in *thyA* and 11 had *deoB* mutated (Figure 6C). We noted as well that evolved D27F strains also carried mutations in other genes, *crr* (7bp deletion), *rpe* and *yhdH* (4bp deletion), however, these were not observed among other trajectories indicating that these are most likely either random passenger mutations or provide only marginal advantage on a specific genetic background. On the other hand, the occurrence of deletions in both *thyA* and *deoB* strongly implies that functional inactivation of these gene products arises as a main mechanism of adaptation to the lack of DHFR function. We reasoned that complementation with either thymidylate synthase or phosphopentomutase should create a significant fitness disadvantage to the *evolved* cells, which can be reverted if these enzymes become inactivated. To verify this prediction, we transformed evolved D27F strains with plasmids expressing either wild type *thyA* or *deoB* and compared the ensuing growth rates of these strains with transformants carrying the same plasmids expressing inactivated *thyA* and *deoB*, respectively. Expression of functional ThyA and especially DeoB were found to be highly

deleterious at low folAmix concentrations, as shown Figure 6 D-E, whereas no significant change in growth was observed upon expressing inactive mutants. The gene *deoB* is part of an operon responsible for deoxyribose degradation, that includes *deoA*, *deoC*, and *deoD*, which is under the control of 2-deoxyribose-5-phosphate-inducible deoR repressor (Hammer-Jespersen and Munch-Ptersen, 1975). DeoB catalyzes the interconversion of 2-Deoxy-D-ribose-1-phosphate and 2-Deoxy-D-ribose-5-phosphate, and, while the former is necessary for synthesis of dTMP from thymine by DeoA, the latter can be further degraded in glycolysis, through conversion into acetaldehyde and D-glyceraldehyde 3-phosphate by DeoC (Figure 6F). Repression of *deo* operon might be deleterious because expression of DeoA is essential for thymine uptake, however co-expressing high levels of DeoB and DeoC can also have a negative effect by directing 2-deoxy-D-ribose-1-phosphate for degradation. Therefore, inactivating *deoB* or *deoC* is an "economy" solution preventing key metabolite in thymine uptake to be wasted in energy production, allowing cells to efficiently use small amounts of thymine available from media supplement for nucleotide synthesis (Dale and Greenberg, 1972). It is also clear that, since we found no mutations in regulatory genes, the origin of the attenuation of the nucleotide-mediated inhibitory/stimulatory effects observed in evolved D27F strain are most likely due to global changes in protein/metabolite concentrations.

Overall, these results show that two key loss of function mutations alone are responsible for the partial adaptation of D27F to loss of DHFR activity and result in significant metabolic, proteomic and regulatory shifts observed in the evolved strains.

**Discussion**

In this study we set to explore the evolutionary mechanisms that follow inactivation of the essential gene that encodes DHFR. Such scenario is akin to antibiotic-mediated target inhibition, but distinct in several aspects. When treated with DHFR inhibitor trimethoprim, bacterial cells recurrently acquire resistance by mutations that decrease drug binding affinity and/or enhance target abundance levels (Toprak et al., 2012). In contrast, genetic inactivation of DHFR at the conditions studied here constrains evolution towards solutions that appear to be mutually exclusive, either restoration of DHFR catalytic activity or adaptation to lack of DHFR function. Interestingly, both evolutionary outcomes were observed. While D27F, D27N and one trajectory in D27G convergently adapted to lack of DHFR function, other trajectories in D27G mutant reached reversion from thymine auxothrophy phenotype. It is important to address the possible reasons for the observed outcome. A major confounding factor could be the presence of background mutations in some of these strains. We note the potential role of the mutation in *upp* gene present in D27G strain. This gene encodes the enzyme uracil phosphoribosyltransferase that participates in pyrimidine salvage pathway and, quite strikingly, appeared consistently depleted in proteomics analysis of all D27 mutants. The slightly higher fitness previously noted for D27G (Figure 2A) could be due some beneficial effect of the *upp* and/or *ymfD* mutation. Conceivably, such small difference in fitness could be a key factor determining the outcome of evolution, but further experiments are needed to test this hypothesis. The codon structure of each mutant studied here and bias in the frequency of each mutation type can also affect the likelihood that phenotype-reverting mutations may arise. We note that the nearest accessible high-fitness Asp/Glu codon for D27F mutant is two nucleotide substitutions away, whereas both D27G and D27N could be rescued by a single G->A or A->G transition, respectively. Only D27G was found to revert, however, not to wild type codon, but to a less catalytically efficient cysteine residue caused a G->T transversion. In this respect, it is rather surprising that D27C mutation was fixed in 3 independent trajectories when transversions are generally regarded to be less likely than transitions. Another important factor in determining the fate of D27 mutant evolution appears to be the early fixation of *thyA* inactivating mutations. Thymidylate synthase is the only known enzyme in *E. coli* able to synthetize dTMP de novo from dUMP, and inactivation of *thyA* inevitably commits the cell to thymine auxothropy; reversion of DHFR function on the *thyA*$^-$ background is not expected to change this phenotype, as it is known that *folA*$^+$ *thyA*$^-$ strains are thymine-dependent (Bertino and Stacey, 1966). Competition between DHFR reversion and *thyA* inactivation thus appears to be decisive in determining the evolutionary solution. However, since inactivation of a gene can be achieved by multiple ways, there is an incomparably greater number of accessible evolutionary trajectories leading to *thyA* knockout when compared with the very limited available mutational routes that restore DHFR function.

Upon *thyA* inactivation, the only available route for dTMP synthesis involves DeoA-catalyzed formation of thymidine from thymine and 2-deoxy-D-ribose-1-phosphate (Figure 6F). However, DeoA is part of the deo operon that is induced by high levels of 2-deoxy-D-ribose-5-phosphate. This regulatory scheme allows cells to recycle excess metabolites into energy production. However, on the background of *thyA* inactivation, the co-expression of *deo* operon proteins creates simultaneously a solution and a problem for the uptake of thymine. Now DeoA becomes an essential protein, but its function is opposed by the roles of DeoC and DeoB which divert 2-Deoxy-D-ribose-1-phosphate into degradation via the glycolysis pathway. The evolutionary solution found in D27 strains converged to the inactivation of a single gene of the operon, *deoB*, an event that is sufficient to block the degradation pathway and yet the regulatory structure is retained. This critical example illustrates the role of regulatory circuits in constraining evolution when disruption of an essential gene forces the system to adopt a metabolic state well beyond normal-operating homeostatic boundaries. Understanding these processes is critical for guidance towards new strategies for antibiotic resistance prevention. Perhaps not surprisingly, we found that pathways of adaptation to genetic inactivation of an antibiotic target provide a route to high levels of resistance. Strikingly, mutations in *thyA* are also known to confer trimethoprim resistance in bacterial clinical isolates of *S. aureus* (Chatterjee et al., 2008; Kriegeskorte et al., 2014) and *H. influenza* (Rodriguez-Arce et al., 2017), which emphasizes the need of laboratory studies of microbial adaptation as useful models to tackle serious global threat.

**Acknowledgements**

This research was supported by NIH grant 5R01GM068670 to E. I. Shakhnovich

**Author Contributions**

Conceptualization, J.V.R and E.I.S, Methodology, J.V.R, and E.I.S, Investigation, J.V.R, Writing – Original Draft, J.V.R and E.I.S, Funding Acquisition, E.I.S

# References

DRAGEN pipeline: http://www.edicogenome.com/dragen_bioit_platform/.

Bader, R., Bamford, R., Zurdo, J., Luisi, B.F., and Dobson, C.M. (2006). Probing the mechanism of amyloidogenesis through a tandem repeat of the PI3-SH3 domain suggests a generic model for protein aggregation and fibril formation. J Mol Biol *356*, 189-208.

Bennett, M.R., Pang, W.L., Ostroff, N.A., Baumgartner, B.L., Nayak, S., Tsimring, L.S., and Hasty, J. (2008). Metabolic gene regulation in a dynamically changing environment. Nature *454*, 1119-1122.

Bershtein, S., Choi, J.M., Bhattacharyya, S., Budnik, B., and Shakhnovich, E. (2015a). Systems-level response to point mutations in a core metabolic enzyme modulates genotype-phenotype relationship. Cell reports *11*, 645-656.

Bershtein, S., Mu, W., Serohijos, A.W., Zhou, J., and Shakhnovich, E.I. (2013). Protein quality control acts on folding intermediates to shape the effects of mutations on organismal fitness. Molecular cell *49*, 133-144.

Bershtein, S., Mu, W., and Shakhnovich, E.I. (2012). Soluble oligomerization provides a beneficial fitness effect on destabilizing mutations. Proc Natl Acad Sci U S A *109*, 4857-4862.

Bershtein, S., Serohijos, A.W., Bhattacharyya, S., Manhart, M., Choi, J.M., Mu, W., Zhou, J., and Shakhnovich, E.I. (2015b). Protein Homeostasis Imposes a Barrier on Functional Integration of Horizontally Transferred Genes in Bacteria. PLoS genetics *11*, e1005612.

Bertino, J.B., and Stacey, K.A. (1966). A suggested mechanism for the selective procedure for isolating thymine-requiring mutants of Escherichia coli. The Biochemical journal *101*, 32C-33C.

Bhattacharyya, S., Bershtein, S., Yan, J., Argun, T., Gilson, A.I., Trauger, S.A., and Shakhnovich, E.I. (2016). Transient protein-protein interactions perturb E. coli metabolome and cause gene dosage toxicity. eLife *5*.

Buescher, J.M., Liebermeister, W., Jules, M., Uhr, M., Muntel, J., Botella, E., Hessling, B., Kleijn, R.J., Le Chat, L., Lecointe, F.*, et al.* (2012). Global Network Reorganization During Dynamic Adaptations of Bacillus subtilis Metabolism. Science *335*, 1099-1103.

Chatterjee, I., Kriegeskorte, A., Fischer, A., Deiwick, S., Theimarm, N., Proctor, R.A., Peters, G., Herrmann, M., and Kahl, B.C. (2008). In vivo mutations of thymidylate synthase (Encoded by thyA) are responsible for thymidine dependency in clinical small-colony variants of Staphylococcus aureus. Journal of Bacteriology *190*, 834-842.

Chen, X., Schulz-Trieglaff, O., Shaw, R., Barnes, B., Schlesinger, F., Kallberg, M., Cox, A.J., Kruglyak, S., and Saunders, C.T. (2016). Manta: rapid detection of structural variants and indels for germline and cancer sequencing applications. Bioinformatics *32*, 1220-1222.

Creek, D.J., Jankevics, A., Breitling, R., Watson, D.G., Barrett, M.P., and Burgess, K.E. (2011). Toward global metabolomics analysis with hydrophilic interaction liquid chromatography-mass spectrometry: improved metabolite identification by retention time prediction. Analytical chemistry *83*, 8703-8710.

Creek, D.J., Jankevics, A., Burgess, K.E., Breitling, R., and Barrett, M.P. (2012). IDEOM: an Excel interface for analysis of LC-MS-based metabolomics data. Bioinformatics *28*, 1048-1049.

D'Souza, G., and Kost, C. (2016). Experimental Evolution of Metabolic Dependency in Bacteria. PLoS genetics *12*, e1006364.

Dale, B.A., and Greenberg, G.R. (1972). Effect of Folic-Acid Analog, Trimethoprim, on Growth, Macromolecular Synthesis, and Incorporation of Exogenous Thymine in Escherichia-Coli. Journal of Bacteriology *110*, 905-+.

Datsenko, K.A., and Wanner, B.L. (2000). One-step inactivation of chromosomal genes in Escherichia coli K-12 using PCR products. Proc Natl Acad Sci U S A *97*, 6640-6645.

Hammer-Jespersen, K., and Munch-Ptersen, A. (1975). Multiple regulation of nucleoside catabolizing enzymes: regulation of the deo operon by the cytR and deoR gene products. Mol Gen Genet *137*, 327-335.

Howell, E.E., Foster, P.G., and Foster, L.M. (1988). Construction of a dihydrofolate reductase-deficient mutant of Escherichia coli by gene replacement. J Bacteriol *170*, 3040-3045.

Johnson, K.A. (2009). Fitting enzyme kinetic data with KinTek Global Kinetic Explorer. Methods Enzymol *467*, 601-626.

Kriegeskorte, A., Block, D., Drescher, M., Windmuller, N., Mellmann, A., Baum, C., Neumann, C., Lore, N.I., Bragonzi, A., Liebau, E.*, et al.* (2014). Inactivation of thyA in Staphylococcus aureus Attenuates Virulence and Has a Strong Impact on Metabolism and Virulence Gene Expression. Mbio *5*.

Krusemann, J.L., Lindner, S.N., Dempfle, M., Widmer, J., Arrivault, S., Debacker, M., He, H., Kubis, A., Chayot, R., Anissimova, M.*, et al.* (2018). Artificial pathway emergence in central metabolism from three recursive phosphoketolase reactions. FEBS J.

Kwon, Y.K., Higgins, M.B., and Rabinowitz, J.D. (2010). Antifolate-induced depletion of intracellular glycine and purines inhibits thymineless death in E. coli. ACS Chem Biol *5*, 787-795.

Liu, G.W., Yong, M.Y.J., Yurieva, M., Srinivasan, K.G., Liu, J., Lim, J.S.Y., Poidinger, M., Wright, G.D., Zolezzi, F., Choi, H.*, et al.* (2015). Gene Essentiality Is a Quantitative Property Linked to Cellular Evolvability. Cell *163*, 1388-1399.

Long, C.P., Gonzalez, J.E., Feist, A.M., Palsson, B.O., and Antoniewicz, M.R. (2018). Dissecting the genetic and metabolic mechanisms of adaptation to the knockout of a major metabolic enzyme in Escherichia coli. Proceedings of the National Academy of Sciences of the United States of America *115*, 222-227.

Lopez-Ibanez, J., Pazos, F., and Chagoyen, M. (2016). MBROLE 2.0-functional enrichment of chemical compounds. Nucleic Acids Res *44*, W201-204.

Luo, H., Gao, F., and Lin, Y. (2015). Evolutionary conservation analysis between the essential and nonessential genes in bacterial genomes. Sci Rep-Uk *5*.

McCloskey, D., Xu, S.B., Sandberg, T.E., Brunk, E., Hefner, Y., Szubin, R., Feist, A.M., and Palsson, B.O. (2018a). Adaptation to the coupling of glycolysis to toxic methylglyoxal production in tpiA deletion strains of Escherichia coli requires synchronized and counterintuitive genetic changes. Metab Eng *48*, 82-93.

McCloskey, D., Xu, S.B., Sandberg, T.E., Brunk, E., Hefner, Y., Szubin, R., Feist, A.M., and Palsson, B.O. (2018b). Adaptive laboratory evolution resolves energy depletion to maintain high aromatic metabolite phenotypes in Escherichia coli strains lacking the Phosphotransferase System. Metab Eng *48*, 233-242.

McCloskey, D., Xu, S.B., Sandberg, T.E., Brunk, E., Hefner, Y., Szubin, R., Feist, A.M., and Palsson, B.O. (2018c). Evolution of gene knockout strains of E-coli reveal regulatory architectures governed by metabolism. Nature communications *9*.

McCloskey, D., Xu, S.B., Sandberg, T.E., Brunk, E., Hefner, Y., Szubin, R., Feist, A.M., and Palsson, B.O. (2018d). Growth Adaptation of gnd and sdhCB Escherichia coli Deletion Strains Diverges From a Similar Initial Perturbation of the Transcriptome. Frontiers in microbiology *9*.

Millard, P., Smallbone, K., and Mendes, P. (2017). Metabolic regulation is sufficient for global and robust coordination of glucose uptake, catabolism, energy production and growth in Escherichia coli. Plos Computational Biology *13*.

Moreb, E.A., Hoover, B., Yaseen, A., Valyasevi, N., Roecker, Z., Menacho-Melgar, R., and Lynch, M.D. (2017). Managing the SOS Response for Enhanced CRISPR-Cas-Based Recombineering in E. coli through Transient Inhibition of Host RecA Activity. Acs Synthetic Biology *6*, 2209-2218.

Pluskal, T., Nakamura, T., Villar-Briones, A., and Yanagida, M. (2010). Metabolic profiling of the fission yeast S. pombe: quantification of compounds under different temperatures and genetic perturbation. Mol Biosyst *6*, 182-198.

Rodrigues, J.V., Bershtein, S., Li, A., Lozovsky, E.R., Hartl, D.L., and Shakhnovich, E.I. (2016). Biophysical principles predict fitness landscapes of drug resistance. Proc Natl Acad Sci U S A.

Rodriguez-Arce, I., Marti, S., Euba, B., Fernandez-Calvet, A., Moleres, J., Lopez-Lopez, N., Barberan, M., Ramos-Vivas, J., Tubau, F., Losa, C.*, et al.* (2017). Inactivation of the Thymidylate Synthase thyA in Non-typeable Haemophilus influenzae Modulates Antibiotic Resistance and Has a Strong Impact on Its Interplay with the Host Airways. Front Cell Infect Mi *7*.

Sangurdekar, D.P., Zhang, Z., and Khodursky, A.B. (2011). The association of DNA damage response and nucleotide level modulation with the antibacterial mechanism of the anti-folate drug trimethoprim. BMC Genomics *12*, 583.

Scheltema, R.A., Jankevics, A., Jansen, R.C., Swertz, M.A., and Breitling, R. (2011). PeakML/mzMatch: a file format, Java library, R library, and tool-chain for mass spectrometry data analysis. Analytical chemistry *83*, 2786-2793.

Tian, J., Woodard, J.C., Whitney, A., and Shakhnovich, E.I. (2015a). Thermal Stabilization of Dihydrofolate Reductase Using Monte Carlo Unfolding Simulations and Its Functional Consequences. Plos Comput Biol *11*.

Tian, J., Woodard, J.C., Whitney, A., and Shakhnovich, E.I. (2015b). Thermal stabilization of dihydrofolate reductase using monte carlo unfolding simulations and its functional consequences. PLoS Comput Biol *11*, e1004207.

Toprak, E., Veres, A., Michel, J.B., Chait, R., Hartl, D.L., and Kishony, R. (2012). Evolutionary paths to antibiotic resistance under dynamically sustained drug selection. Nature genetics *44*, 101-105.

Toprak, E., Veres, A., Yildiz, S., Pedraza, J.M., Chait, R., Paulsson, J., and Kishony, R. (2013). Building a morbidostat: an automated continuous-culture device for studying bacterial drug resistance under dynamically sustained drug inhibition. Nature protocols *8*, 555-567.

**STAR Methods**

**Contact for Reagent and Resource Sharing**

Further information and requests for resources and reagents should be directed to and will be fulfilled by the Lead Contact, Eugene I. Shakhnovich (shakhnovich@chemistry.harvard.edu).

**Experimental Model and Subject Details**

*Construction of D27 mutants*

Inactive D27 mutants were created by lambda-red recombination (Datsenko and Wanner, 2000) according to a previously described procedure (Bershtein et al., 2012) with some modifications. Briefly, a pKD13 plasmid was modified to contain the entire regulatory and coding sequence of folA gene, flanked by two different antibiotic markers (genes encoding kanamycin (kanR) and chloramphenicol (cmR) resistances) and approximately 1kb homologous region of both upstream and downstream chromosomal genes flanking folA gene (kefC and apaH, respectively). The entire cassette was amplified and transformed into BW25113 cells with induced Red helper plasmid pKD46, and cells were recovered in SOC medium containing 1x folAmix (adenine 20 ug/mL, inosine 80 ug/mL, thymine 200 ug/mL, methionine 20 ug/mL and glycine 20 ug/mL). Transformants were plated in agar media containing both antibiotics and folAmix. Correct integration of the desired mutations was confirmed by Sanger sequencing of *folA* gene. Plasmid pKD46 was removed by plating cells at 37 °C twice in the absence of antibiotic selection.

**Method Details**

*Automated experimental evolution*

A liquid handling instrument Tecan Freedom Evo 150 equipped with Magellan plate reader and Liconic shaker was used in this work. The experiments were done at 30 ˚C and using M9 minimal media supplemented with 2g/L glucose, 34µg/mL chloramphenicol and 50µg/mL kanamycin. The general procedure involves up to four 96-well plates that are used for serial dilutions of bacterial cultures. Each working plate can carry 8 trajectories that are positioned in a single column. In this work we used the first and eight wells in the column with media alone to control for contamination. The experiment starts by placing the 200 µL cultures/control in the first column of the 96-well plate. All the four plates are incubated in a shaker at 30 ˚C and at every 30 min the OD of each plate is determined alternately. The growth rate of each culture is calculated from OD measurements over time. To ensure proper comparison, the growth rate was computed only from OD values within a specified range (0.1-0.25). When the average OD of the six experimental replicates in a plate exceeds a threshold of 0.30, each culture is diluted into the next adjacent column by mixing a calculated volume of both culture and fresh medium and folAmix so that the initial OD is 0.01 in a total of 200 µL. At this point, the remaining portion of the previous culture is mixed with glycerol in an auxiliary plate and subsequent freezed at -80C. The cycle is repeated through the entire experiment. At every passage, the growth rate is compared with a threshold value (0.16 h$^{-1}$) and whenever a culture exceeds this value the concentration of folAmix is halved.

*Media and Growth Conditions*

Cell cultures were recovered by inoculating fresh M9 media medium supplemented with 0.2% glucose and 1x folAmix with a portion of -80°C glycerol stocks taken at different passages. After recovery for several hours, the cultures were plated in selective agar media containing 1x folAmix and incubated at 30°C. Individual colonies were grown in M9-media+folA mix and stored in glycerol at -80°C for later analysis. Cell cultures grown overnight were diluted in fresh M9 minimal media containing 1x folAmix and antibiotics and were grown for additional 4-6 h. Cultures were then pelleted by centrifugation and washed 3 times in fresh M9 without folAmix. Microplates containing 150 µL of M9 minimal with 0.8g/L glucose, antibiotics, and varying concentrations of folAmix were then inoculated with each culture at a starting OD of 0.0005. Growth measurements were performed in a Infinite® 200 PRO plate reader for 48h at 30°C with constant shaking. Growth rate values are represented as mean ± SD from at least 3 biological replicates.

*Protein purification and characterization*

D27 mutant fused to C-terminal (6x) His-tag were overexpressed using pFLAG expression vector under isopropyl β-D-1-thiogalactopyranoside (IPTG) inducible T7 promoter. The recombinant proteins were purified from lysates on Ni-NTA columns (Qiagen) as described previously(Rodrigues et al., 2016). DHFR kinetic parameters were derived from the analysis of progress-curve kinetics of NADPH oxidation in the presence of dihydrofolate using the software Kintek Explorer (Johnson, 2009) as described before(Rodrigues et al., 2016).

*Predicting fitness of D27 mutants*

The fitness of E. coli DHFR mutants can be predicted from in vitro biophysical parameters using a simple metabolic model described in an earlier work (Rodrigues et al., 2016). The metabolic flux through DHFR reaction in DHFR mutant strains, normalized to wild type, can be approximated to:

$$V_{dhfr}^{norm} = \frac{\frac{k_{cat}^{mut}}{K_M^{mut}} \cdot [DHFR]^{mut}}{\frac{k_{cat}^{EcoliWT}}{K_M^{EcoliWT}} \cdot [DHFR]^{WT}} \cdot \frac{1}{\left(1 + \frac{\alpha \cdot [TMP]_{medium}}{K_i^{mut}}\right)} \quad \text{eq. 2}$$

Where, $\frac{k_{cat}}{K_M}$, $K_i$ and $[DHFR]$ are, respectively, the catalytic efficiency, trimethoprim inhibition constant and intracellular protein abundance of the mutant or of the wild type, and $[TMP]_{medium}$ and $\alpha$ are, respectively, the concentration of trimethoprim in the culture medium and the ratio of intracellular vs extracellular concentrations of trimethoprim, defined previously to be 0.1

(Rodrigues et al., 2016). To connect flux to fitness, we first measure the growth rate of wild type *E. coli* cells at various concentrations of DHFR inhibitor trimethoprim.

Then fitness is plotted against the flux through DHFR reaction calculated at each concentration of inhibitor using equation 2, and catalytic parameters determined in vitro (Table 1). Protein abundance is determined by measuring the total catalytic activity in cell lysates, as described earlier (Rodrigues et al., 2016). Finally, we use equation 3:

$$Gr^{norm} \sim \frac{V_{dhfr}^{norm}}{\left(B + V_{dhfr}^{norm}\right)} \quad \text{eq. 3}$$

to fit the fitness vs $V^{norm}$ data points to determine the parameter *B*, which represents the normalized flux at which growth rate is reduced by half. From this equation, the in vitro parameters determined for D27 mutants are used in equation 2 and 3 to predict fitness. The protein abundance of D27 mutants, however, cannot be determined from enzymatic assays in cell lysates, as these are inactive variants. Instead, protein abundance was predicted using an inverse relationship with relative bis-ANS fluorescence of those variants (Bershtein et al., 2013; Rodrigues et al., 2016). In qualitative terms, our model accurately predicted that the D27C variant is able to grow in the absence of folAmix supplement. However, the measured fitness was somewhat higher than the value predicted by the calculations. It is possible that some assumptions are not entirely met. One important assumption is that the concentration of DHFR and substrate dihydrofolate are either always constant or change only as a function of the flux calculated by equation S1. The latter implies that, for every given value of flux, the changes in DHFR and dihydrofolate concentration will always be the same, and thus always comparable regardless of what parameters are used as input in equation S1. However, this might not be always true. For example, a tight feedback control regulates the DHFR promoter activity, in response to the metabolic needs of the cell (Bershtein et al., 2015a; Bershtein et al., 2015b). Likewise, a decrease in DHFR catalytic function is also expected to result in a considerable level of substrate build up. In these conditions, the magnitude of substrate $K_M$ for each variant might become more relevant. In our calculations, however, substrate saturation is not considered. This may lead to deviations, especially when KM values differ significantly, as we find in this work. How much these dynamic processes may differently affect the flux through DHFR reaction in each mutant is still to be determined.

*Effect of deoB and thyA expression*

The genes *deoB* and *thyA* and corresponding mutants *deoB* p.309Stop and *thyA* 755-762del were cloned into a modified pTRC plasmid in which the *laqI* and promoter region were replaced by the *tetR* repressor gene and promotor region derived from plasmid Plasmid pEM-Cas9HF1-recA56 (addgene Addgene plasmid # 89962 (Moreb et al., 2017)). Transformed strain cells were grown in

M9 minimal media + folAmix, then washed with fresh M9 media without folAmix and plated at different concentration of folAmix. Growth rate values are the mean ± SD of at least 3 biological replicates.

*Metabolite extraction and LC-MS analysis*

Cultures of mutant strains (naïve and evolved D27F) and wild type (30 mL) were grown in parallel in M9 minimal media supplemented with 2 g/L glucose in a 250 mL flask at 30C. At an OD of 0.2-0.25 the cultures were centrifuged. The volume of culture to be pelleted was calculated so that the product OD × volume of cell culture (mL) = 5. The pellet was mixed with 300 µl of 80:20 ratio of methanol:water that had been pre-chilled on dry ice. Samples were vortexed and incubated in dry ice for 10 min followed by centrifugation at 4˚C for 10 min at maximum speed. The supernatant was collected, and the pellet was processed by repeating the procedure. Samples were stored at -80˚C until analyzed by mass spectrometry. At least three independent biological replicates were analyzed for each strain. LC-MS analysis in the positive and negative mode was performed as previously described (Bhattacharyya et al., 2016). Retention times of several metabolites of the nucleotide biosynthesis pathway were determined from the analysis of pure compounds.

*LC/MS TMT Proteomics*

Cell cultures from isolated colonies were grown at 30˚C in M9 minimal media supplemented with 0.2% glucose and were collected by centrifugation during mid-exponential phase (OD ~ 0.2-0.25), well below saturation due to oxygen limitation (typically at OD>0.8-0.9). Global proteomic analysis was performed as described previously (Bershtein et al., 2015a).

*Whole-genome sequencing*

Sequencing was performed on isolated colonies on Illumina MiSeq in 2x150 bp paired-end configuration (Genewiz, Inc., South Plainfield, NJ). The raw data were processed with the DRAGEN pipeline on default settings, using the E. coli K-12 substr. MG1655 reference genome (GenBank accession no. NC_000913.3). After the alignment completed, structural variations (SV) including large INDELs (above 50) were detected using the program Manta(Chen et al., 2016).

**Quantification and Statistical Analysis**

*Metabolomics data analysis.*

Data analysis was performed using the software packages MzMatch (Scheltema et al., 2011) and IDEOM (Creek et al., 2012) for untargeted analysis. For peak assignment in untargeted analysis, IDEOM includes both peak m/z values and predicted retention times calculated based on chemical descriptors (Creek et al., 2011). A list of 32 experimentally measured retention times was initially used to calibrate retention time predictions. The retention time of putatively identified metabolites were found to correlate fairly well with the values included in IDEOM ($R^2$=0.73) and published in other studies (Pluskal et al., 2010) (Pluskal et al., 2010) ($R^2$=0.88 and $R^2$=0.61, respectively). For this reason, additional metabolites from those sources that closely matched IDEOM assignments were treated as standards in the identification routine. Following identification, Z-scores with respect to wild type ($Z^{D27F\_naïve/WT}$) or to naïve D27F strain ($Z^{D27F\_evo/D27F\_naïve}$) were calculated using equation 1. Z-scores determined for each set *i* of *n* biological replicates were combined for each metabolite *met* using the expression $z^{met}_{combined} = \frac{\sum_{i}^{n} z_i^{met}}{\sqrt{n}}$. The set of metabolites with highest absolute combined Z-scores (>1.96) was selected to perform an overrepresentation (enrichment) analysis of categorical annotations using MBrole2.0 (Lopez-Ibanez et al., 2016), using as reference the metabolic pathways of *E. coli* from KEGG database.

*Proteomics data analysis*

Z-scores were computed using equation 1. For grouping analysis, we used the functional and regulatory classification group sets as (Sangurdekar et al., 2011). For each set of genes belonging to a group we employed a one-sample t-test which provides the p-value against the null hypothesis that the group of genes was drawn from a normal distribution and considered that a given group of genes is upregulated or downregulated with respect to a reference if the average of Z-scores of that group is positive or negative, respectively.

**Table S1-Primers list**

| tetR_pTRC_for | ACACCATCGAATGATATCGACGTCTTAAGACCCACTTT |
|---|---|
| tetR_thyA_rev | CTAAATACTGTTTCATAGATCCGAAGTCCTCTTTAGATC |
| tetR_thyA_for | GACTTCGGATCTATGAAACAGTATTTAGAACTGATGC |
| tetR_thyA-rev | GACTTCGGATCTATGAAACAGTATTTAGAACTGATGC |
| tetR_deoB_for | GACTTCGGATCTATGAAACGTGCATTTATTATGGTGC |
| pTRC_deoB_rev | AAAACAGCCAAGCTTTCAGAACATGGCTTTGCCATATTCC |
| deoB_pTRC_for | AAAGCCATGTTCTGAAAGCTTGGCTGTTTTGGCGGATGAG |
| deoB_tetR_rev | AAATGCACGTTTCATAGATCCGAAGTCCTCTTTAGATC |
| CapR-Chrom-Flanking rev | TTAGGATGAGGGTTTCGTTTCCGGTTCATC |
| PCRseq_KefC_for2 | CTGCTCGGTTTCCTCATCATCAA |
| P4_chrom_flanking for | TTTTGTATAGAATTTACGGCTAGCGCCG |
| PCRseq_apaH_rev | CGTCCCTTTCAGCATCGACATT |
| pKD13_post_downsream_for | CTTATCACTGATCAGTGAATTAATGGCG |
| pKD13_post_upstream_rev | GACAATAACCCTGATAAATGCTTCAATAATATTG |
| Upstream_capR_for | GAAGAAGGTAAACATACCGGCAACATGGCGGATGAACCGGAAACGAAACCCTCATCCTAATCATGATCATCGCAGTACTGTTG |
| P4downstream_rev | AAGGCCGGATAAGACGCGACCGGCGTCGCATCCGGCGCTAGCCGTAAATTCTATACAAAACTGTCAAACATGAGAATTAATTC |
| Ampl_RRfolA_for | GTGCCGATCAACGTCTCATTTTCG |
| Ampl_RRfolA_rev | GCTTCCTCGTGCTTTACGGTATCG |
| PCRseq_RRfolA_rev | GCCTTCTATCGCCTTCTTGACGA |
| D27Fmut_For | TTTCTCGCCTGGTTTAAACGCAACACCTTAAATAAAC |
| D27Gmut_for | GGCCTCGCCTGGTTTAAACGCAACACCTTAAATAAA |
| D27Nmut_for | AATCTCGCCTGGTTTAAACGCAACACCTTAAATAAAC |
| D27_rev | GGCAGGCAGGTTCCACGGCATGG |

**Excel Tables**

LC/MS TMT Proteomics Data: Proteomics_data.xlsx

Sequencing data: Sequencing_data.xlsx

Metabolomics data: Metabolomics_data.xlsx